\documentclass[usenatbib,onecolumn]{mn2e}

\usepackage{epsfig}

\usepackage{longtable}
\usepackage{amsmath}

\graphicspath{{./figures/}}

\def\be		{\begin{equation}}
\def\ee		{\end{equation}}
\def\ba		{\begin{aligned}}
\def\ea		{\end{aligned}}

\begin{document}

\title{Feasibility of observing Hanbury Brown and Twiss phase}

\author[T. Wentz and P. Saha]{Tina Wentz$^1$ and Prasenjit Saha\\
Physik-Institut, University of Zurich, Winterthurerstrasse 190, 8057 Zurich, Switzerland}

\maketitle

\footnotetext{$^1$Present address: TNU, Institute for Biomedical
  Engineering, University of Zurich and ETH Zurich, Wilfriedstrasse 6, 8032 Zurich,
  Switzerland.}

\begin{abstract}
The interferometers of Hanbury Brown and collaborators in the 1950s and 60s, and their modern descendants now being developed (intensity interferometers) measure the spatial power spectrum of the source from intensity correlations at two points. The quantum optical theory of the Hanbury Brown and Twiss (HBT) effect shows that more is possible, in particular the phase information can be recovered by correlating intensities at three points (bispectrum). In this paper we argue that such 3 point measurements are possible for bright stars such as Sirius and Betelgeuse using off the shelf single photon counters with collecting areas of the order of 100m$^2$. It seems possible to map individual features on the stellar surface. Simple diameter measurements would be possible with amateur class telescopes.
\end{abstract}

\section{Introduction}

The Hanbury Brown and Twiss (HBT) effect can be thought of as Young's double slit experiment using narrowband but incoherent light. Since the source is incoherent the phase is fluctuating \citep[see for example Figure 7.1 in][]{2006iai..book.....L} and there is no consistent phase difference between the two waves emitted at the slits and thus there will be no interference patterns produced by constructive and destructive superposition visible on the screen, the time averaged intensities (the complex conjugate products of the incoming electric fields) coming from both slits will just add up. But due to the random fluctuations of the phases the intensities will fluctuate. Counting individual photons at photons at two different points on the screen will result in a coincidence rate slightly higher than the Poisson coincidence rate, the intensity fluctuations are correlated. This additional term is what is called the HBT effect.

In astronomy we need to replace the slits with a two-dimensional superposition of sub-sources. Integrating the amplitude contributions over the new, extended source amounts to taking the Fourier transform of the intensity or brightness distribution on the sky (Van Cittert Zernike theorem). The normalized spatial Fourier transform is called the complex visibility $V$ and can, in principle, be measured at different spatial distances or baselines.  But measuring visibility directly is very difficult for astronomical sources because it requires maintaining optical quality mirrors over long baselines to sub wavelength accuracy. Alternatively, information on the visibility can be obtained by considering the normalized intensity correlation (the probability of coincident detection at two detectors) for two detectors ($1$ and $2$):
\be
C_{12} = 1 + |V_{12}|^2
\label{eq:corr}
\ee
This equation is valid over a coherence time which is the time scale
\be
\Delta\tau \sim 1/\Delta\nu = \frac{\lambda^2}{c \,\Delta\lambda}
\ee
on which the phases of incoherent light fluctuate. The phase information of the electrical field is lost when measuring the amplitude squared.  The time resolution $\Delta t$ for measuring intensity correlations or photon coincidences should be as quick as possible.  If $\Delta t$ is shorter than $\Delta\tau$, the full benefit of the signal in equation \eqref{eq:corr} is obtained.  If $\Delta t\gg\Delta\tau$ the signal to noise per measurement interval ${\rm SNR}(\Delta t)\ll1$.  Nevertheless, sufficient SNR can be built up by collecting data over many many $\Delta t$.  The great advantage of HBT interferometry is that optical parts need to be accurate only to $\ll c\Delta t$.  For the $\Delta t$ achievable nowadays, optical-path tolerances of a millimetre would be adequate.  In contrast, standard interferometry requires optical paths to be precise to better than a wavelength.

\begin{figure}
\begin{minipage}[b]{0.5\linewidth}
\centering
\includegraphics[scale=.45]{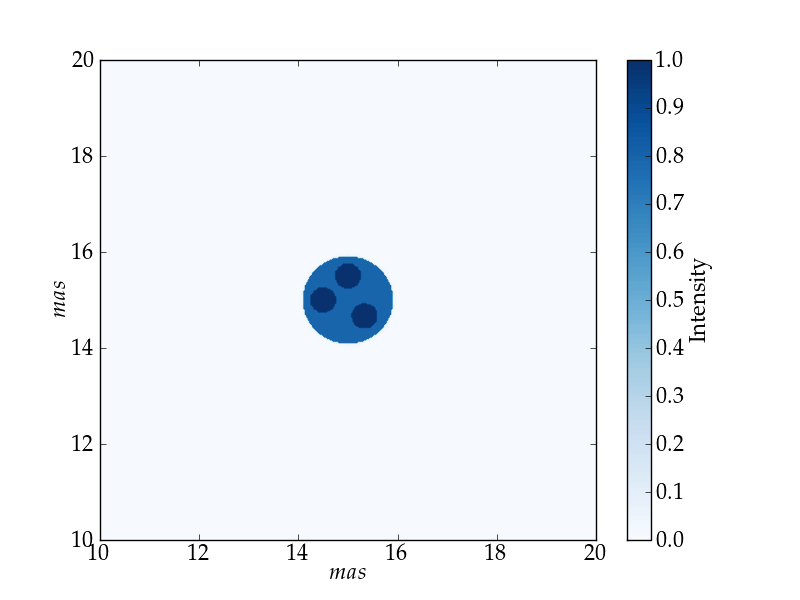}
\end{minipage}
\hspace{0.5cm}
\begin{minipage}[b]{0.5\linewidth}
\centering
\includegraphics[scale=.45]{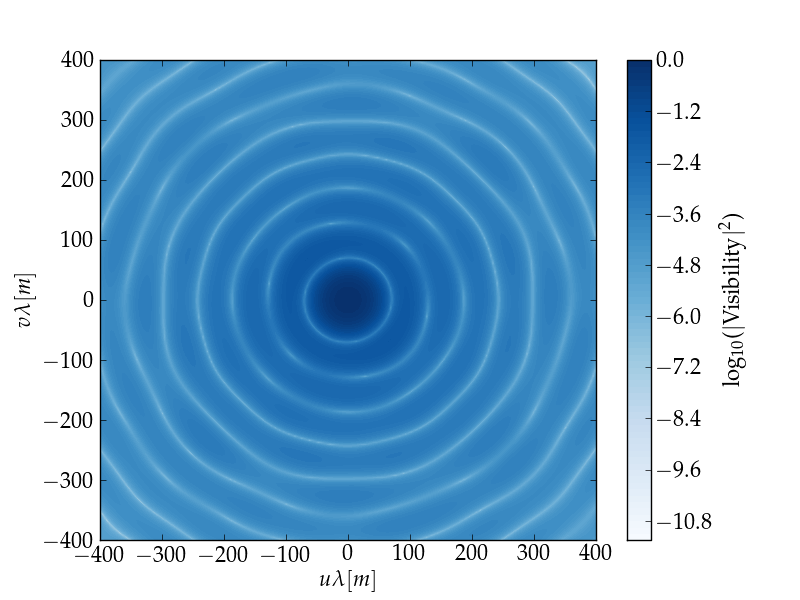}
\end{minipage}
\begin{minipage}[b]{0.5\linewidth}
\centering
\includegraphics[scale=.45]{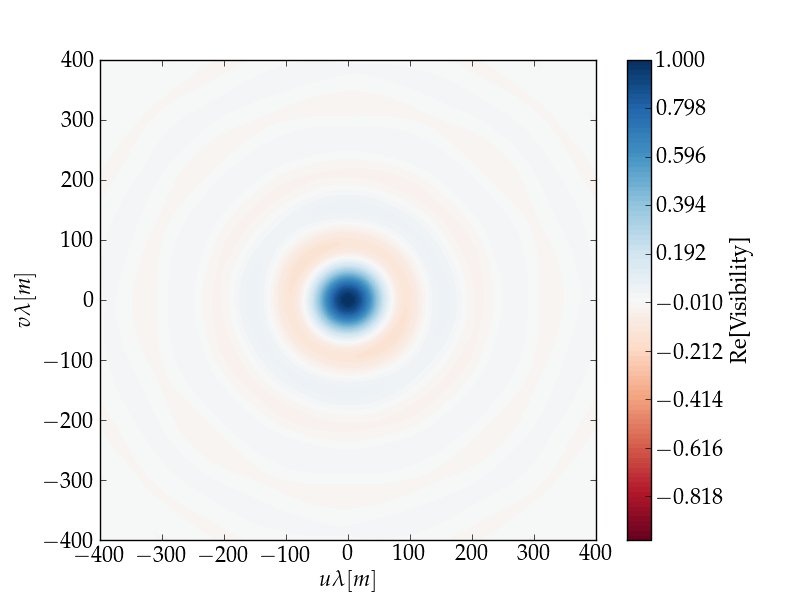}
\end{minipage}
\hspace{0.5cm}
\begin{minipage}[b]{0.5\linewidth}
\centering
\includegraphics[scale=.45]{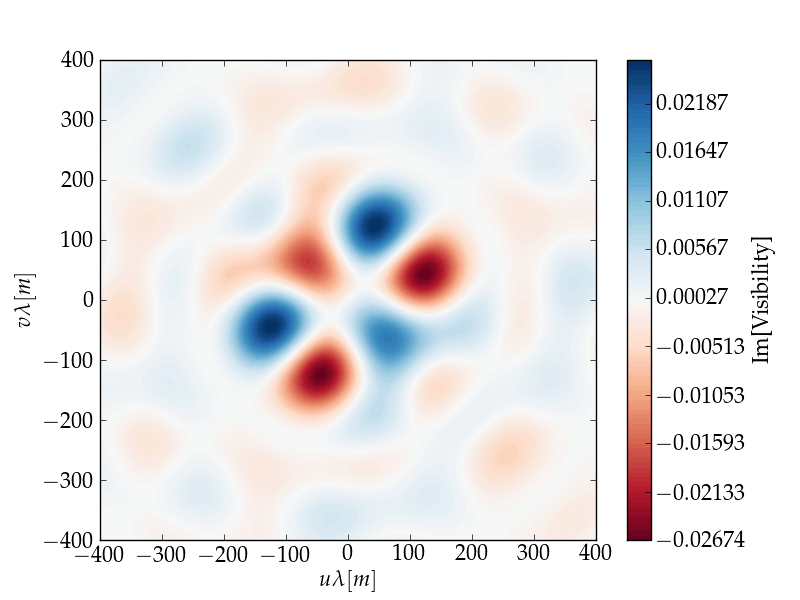}
\end{minipage}
\caption{Simulation of a structured, crudely limb-darkened disc of 1~mas radius (inspired by Betelgeuse): Top left shows the simulated source brightness distribution, top right shows the absolute value of the visibility squared, the 2 point correlation term in equation (\ref{eq:corr}) as simulated using a 2D FFT algorithm. The axes of this and the bottom panels are baselines in meters, for $\lambda=500\rm\,nm$. For completeness the bottom panels show the real and imaginary parts of the complex visibility. Note that the imaginary part goes to zero for point symmetric sources.}\label{fit}  
\label{fig:simone}
\end{figure}

In \cite{1958RSPSA.248..222B} famously implemented these ideas in the first stellar intensity interferometer, which measured the diameter of Sirius.  In the 1960s the Narrabri intensity interferometer was built to measure more stellar diameters \citep{1974MNRAS.167..121H}. At that time intensity interferometry was limited to blue-sensitive counting equipment, and after the Narrabri instrument had observed all the hot stars it could, the technique was abandoned for a long time. The possibilities of new, faster counters in red and infrared ranges have recently brought intensity interferometry back into focus and arrays of telescopes such as the proposed Cherenkov Telescope Array (CTA) will allow us to exploit the effect to a greater extent in the years to come \citep[see for example][]{2013APh....43..331D}. Some illustrative examples are shown in Figures \ref{fig:simone} and \ref{fig:fftld}.

The loss of phase information, as evident from the absolute value in equation (\ref{eq:corr}) is the major shortcoming of intensity interferometry. But it is not a fundamental shortcoming, because from the theory of quantum optics developed in the 1960s \citep[see][for a review]{glauber2006}, phase information can be obtained by using three telescopes in concert.

The normalized three point correlation is:
\be
C_{123} = 1 + |V_{12}|^2 + |V_{23}|^2 + |V_{31}|^2 + 2{\rm Re}[V_{12}V_{23}V_{31}] .
\label{eq:3corr}
\ee
The product in the last term is the real part of the spatial bispectrum of the source.  Its phase is well known in radio astronomy as the closure phase, and useful for eliminating local atmospheric and other influences on phase measurements for individual detectors in a three-antenna setup.  The term has also long been studied in the context of diverse laboratory experiments: \cite{1963JAP....34..875G} already used photon triple-coincidence as a tool for spectral measurement; \cite{1978ApOpt..17.2047S} constructed an imaging system using three-point HBT; \cite{1983JAP....54..473F} performed some remarkable experiments showing three- and four-point acoustic HBT.  The analyses in these works is in terms of classical waves and intensities, so it does not necessarily apply to photon counting, but fortunately it turns out \citep[see for example][]{1964PPS....84..435M} that for ordinary light, quantum fields can be replaced with classical and intensities interpreted as photon-detection probabilities.

\cite{2014MNRAS.437..798M} recently reviewed the theory of higher order correlations and provided a simple way of estimating the achievable SNR. In the present work, we will study the feasibility of detecting three-point HBT for astronomical sources. In the following sections we will simulate signals for two and three point correlation measurements of different, simple sources and make some estimates about possible SNR. We will not consider the problem of reconstruction of the actual phase but algorithms for related problems are known \citep[see, for example][]{Kang:1991:529}.

Let us now briefly preview our results.

In the next Section we show simulations of the complex visibilities and resulting correlation signals. For example, Figure~\ref{fig:simone} shows a simulation of a structured disc, inspired by reconstructions of Betelgeuse. Comparing that to the other figures in Section~\ref{sec:cv} we can immediately see that adequate $(u,v)$ coverage allows to see differences in shape and structure. We also plot the 3-point correlation signals (specifically the bispectra) for a given baseline for different sources.

In Section~\ref{snrchap} we rederive the well-known but counterintuitive result that for two-point correlation, the SNR is independent of bandwidth, meaning that decreasing the bandwidth and hence decreasing the count rates will not change the SNR for HBT detection.  A related feature is that the observation time needed goes as the inverse square of the collection area. With larger telescopes such as H.E.S.S. (12~m diameter dishes) and the planned CTA (7m, 12m, and 23m are proposed) this will allow for significant measurements on very short timescales. We also want to emphasize again the possibilities in using arrays with many telescopes increasing the possible $(u,v)$ \citep[see for example][]{2013APh....43..331D}. For the brightest stars, size measurements would be possible with even a 10~cm diameter mirror with modern off-the-shelf single-photon correlators.  With a 10 cm diameter telescope and a counting system at 50\% efficiency one could even achieve an SNR of 1 in 36 seconds integration time for Capella b for example. The small baselines possible with such a setup should allow the detection of the 2 point correlation for this binary system, that would require a about a 2 meter aperture to be resolved by a single telescope.

While we have considered nearby stars as examples, the results can be trivially scaled.  For example, scaling Sirius up in size and distance by $10^4$ and scaling in luminosity by $10^8$ (keeping apparent size, luminosity and effective temperature the same) would be not unlike a supernova in the LMC.  This suggests that supernova shells in nearby galaxies could be rather easily resolved --- the equivalent of SN~1987a with very modest equipment and the equivalent of SN~2011fe in M101 with the sort of instruments now being proposed.

Section~\ref{snrchap} also shows the necessary observation times to reach a signal to noise ratio of one for the three-point correlation signal. Skipping ahead to Figure \ref{fig:qtobs3} we can immediately see that measuring the bispectrum in the visible range using off the shelf photon counters is easily achievable for larger telescopes as the observation time goes with the inverse cubed of the collection area. To emphasize this point: measuring the three-point HBT signal will be feasible and informative for CTA type mirrors.

\label{sec:cv}

\section{Complex visibility vs HBT observables}

\begin{figure}
\begin{minipage}[b]{0.5\linewidth}
\centering
\includegraphics[scale=.45]{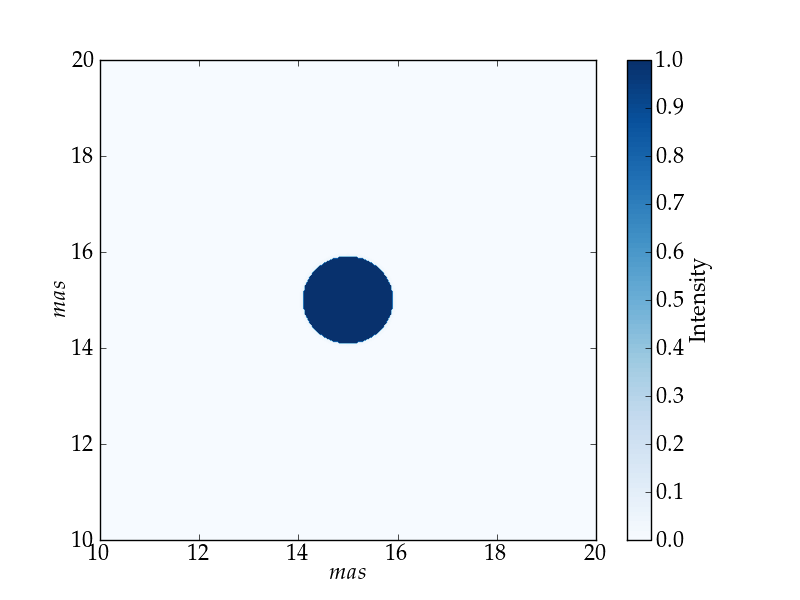}
\end{minipage}
\hspace{0.5cm}
\begin{minipage}[b]{0.5\linewidth}
\centering
\includegraphics[scale=.45]{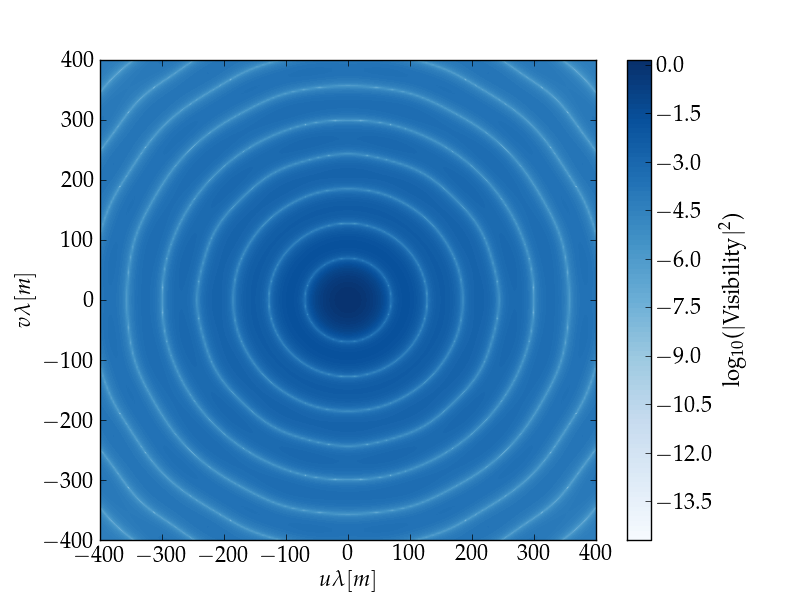}
\end{minipage}
\begin{minipage}[b]{0.5\linewidth}
\centering
\includegraphics[scale=.45]{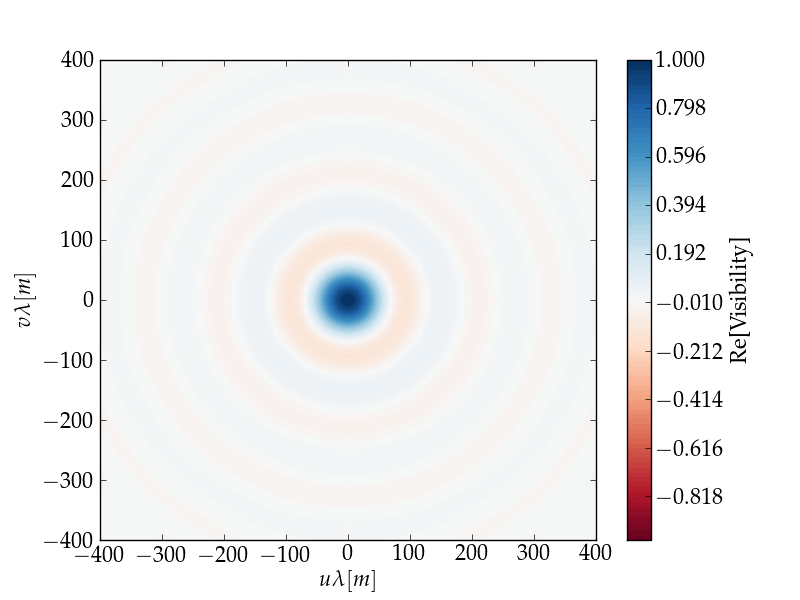}
\end{minipage}
\hspace{0.5cm}
\begin{minipage}[b]{0.5\linewidth}
\centering
\includegraphics[scale=.45]{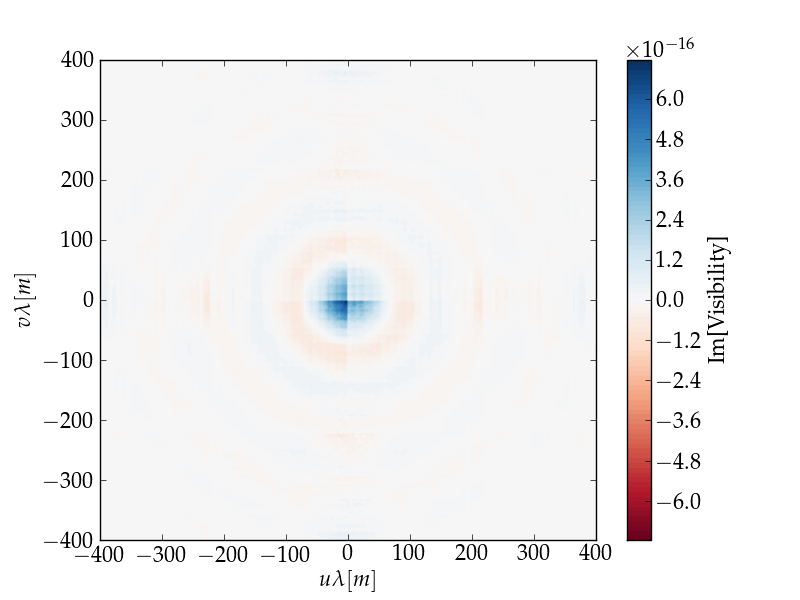}
\end{minipage}
\begin{minipage}[b]{0.5\linewidth}
\centering
\includegraphics[scale=.45]{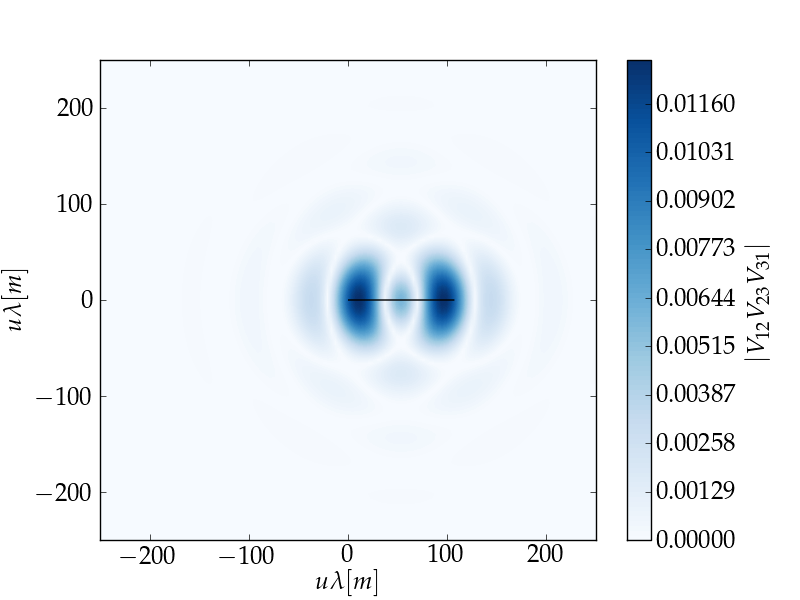}
\end{minipage}
\hspace{0.5cm}
\begin{minipage}[b]{0.5\linewidth}
\centering
\includegraphics[scale=.45]{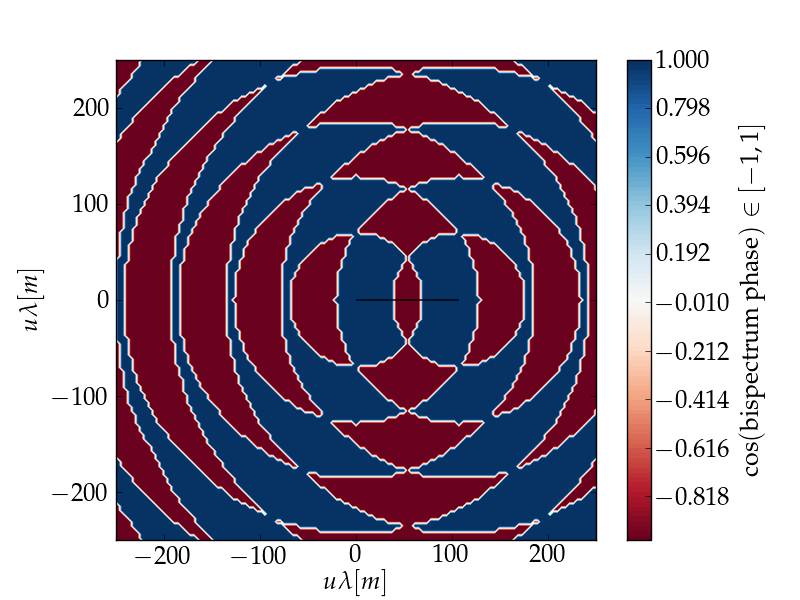}
\end{minipage}
\caption{Simulation of a roughly limb darkened disc of $1\rm\,mas$ radius: left the simulated source, on the right the squared visibility. As we can see the imaginary part of the complex visibility is zero except for roundoff errors. The two lower panels show absolute value and cosine of the phase of the bispectrum at the position of a third detector with respect to a fixed 107~m long baseline of two detectors as indicated by the black line. Note that these quantities cannot be measured directly but their product can.}
\label{fig:fftld}
\end{figure}

The Van Cittert Zernike theorem allows us to simulate complex visibilities using a 2D FFT. Figure~\ref{fig:simone} shows a simulation of a structured, crudely limbdarkened disc of 1~mas radius.\footnote{The baseline scales are always in metres: units of frequency space $(u,v)$ multiplied by an assumed measurement wavelength $\lambda=500\rm\,nm$.} It features three spots that are 25\% brighter than the rest of the disc. This mimics the features observed on Betelgeuse using standard optical interferometry \citep{2000MNRAS.315..635Y,2009A&A...508..923H}. Figure \ref{fig:fftld} shows a similar crudely limb darkened disc without the spots. The power spectrum shows extremely subtle differences, though, looking at real and imaginary parts it is clear one brightness distribution is symmetrical and the other one is not. Clearly it is desirable to have information about the phase of the complex visibility.

\begin{figure}
\centering
\includegraphics[width=1.\linewidth]{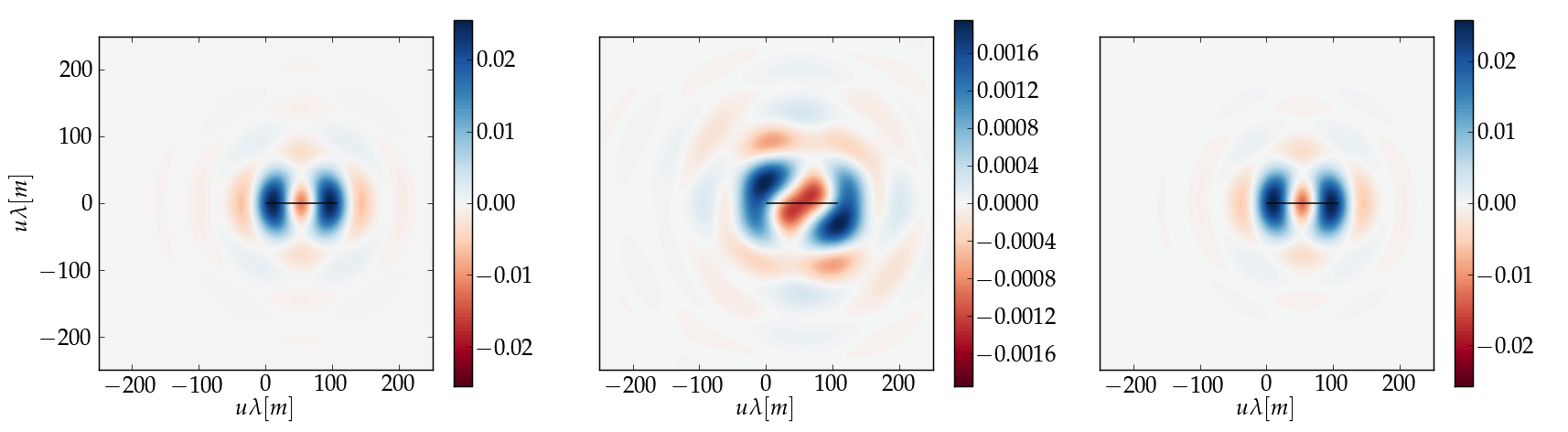}
\caption{Bispectra $2{\rm Re}[V_{12}V_{23}V_{31}]$ as the position of one detector is varied while the other two detectors are fixed (at opposite ends of the black line indicating a baseline of 107~m as before). The left panel corresponds to the source in figure \ref{fig:fftld} while the right panel to the disc in figure \ref{fig:simone}. The center panel shows the difference.}
\label{fig:diffat}
\end{figure}

The bispectrum is a function of two baselines, so it cannot be shown in a single figure. In Figure \ref{fig:diffat} we show an example with two detectors (hence one baseline) fixed. \footnote{A program to simulate other baselines as well as different sources, such as a binary, is provided in the supplementary material.}

Features of 25\% difference in brightness appear at the level of $10^{-3}$ in the bispectrum. To assess the feasibility of measuring this one has to calculate the signal to noise ratio.

\section{Count rates and Signal to Noise}
\label{snrchap}

\subsection{count rates}

\begin{figure}
\begin{minipage}[b]{0.5\linewidth}
\centering
\includegraphics[scale=.45]{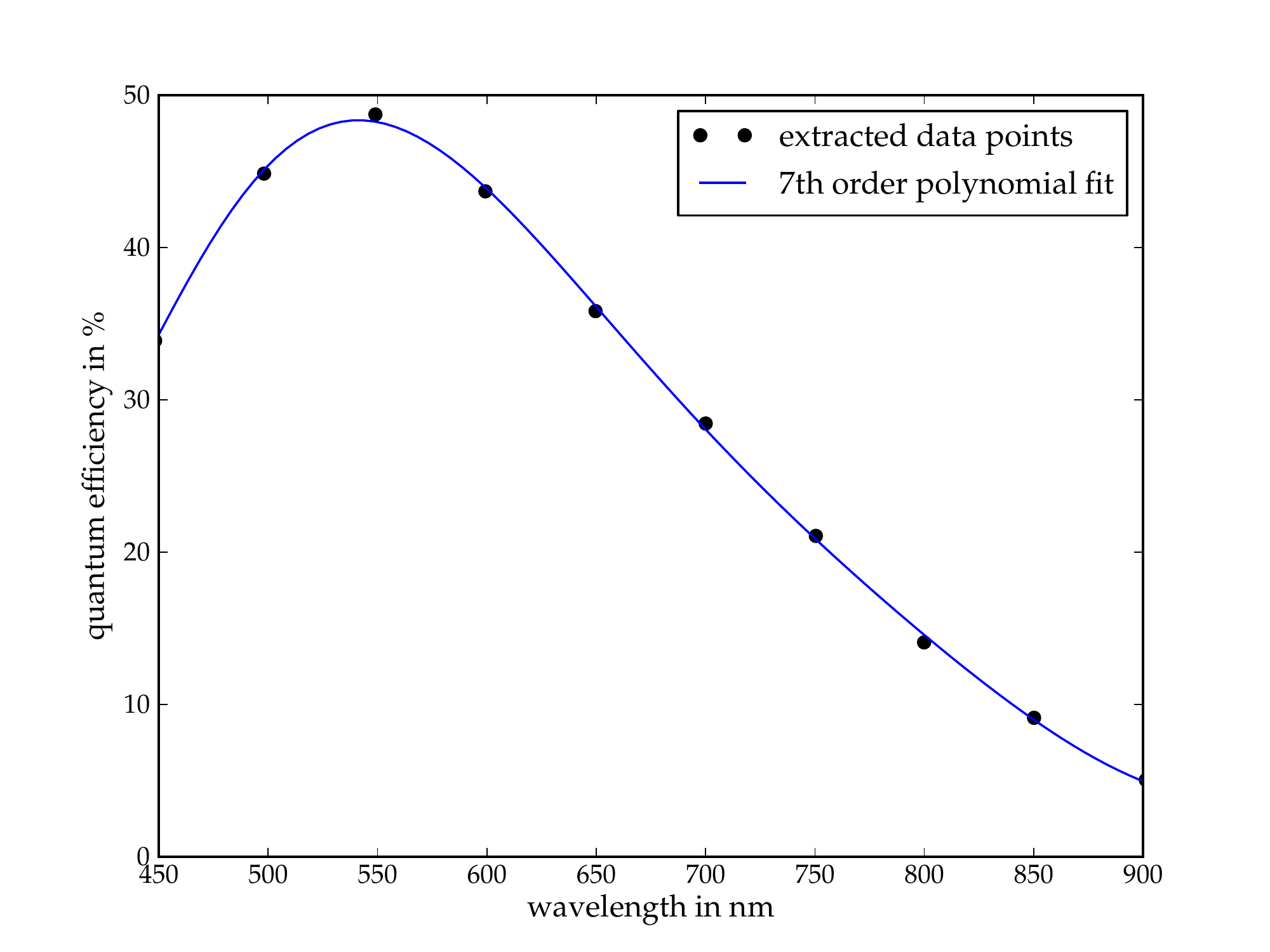}
\end{minipage}
\hspace{0.5cm}
\begin{minipage}[b]{0.5\linewidth}
\centering
\includegraphics[scale=.45]{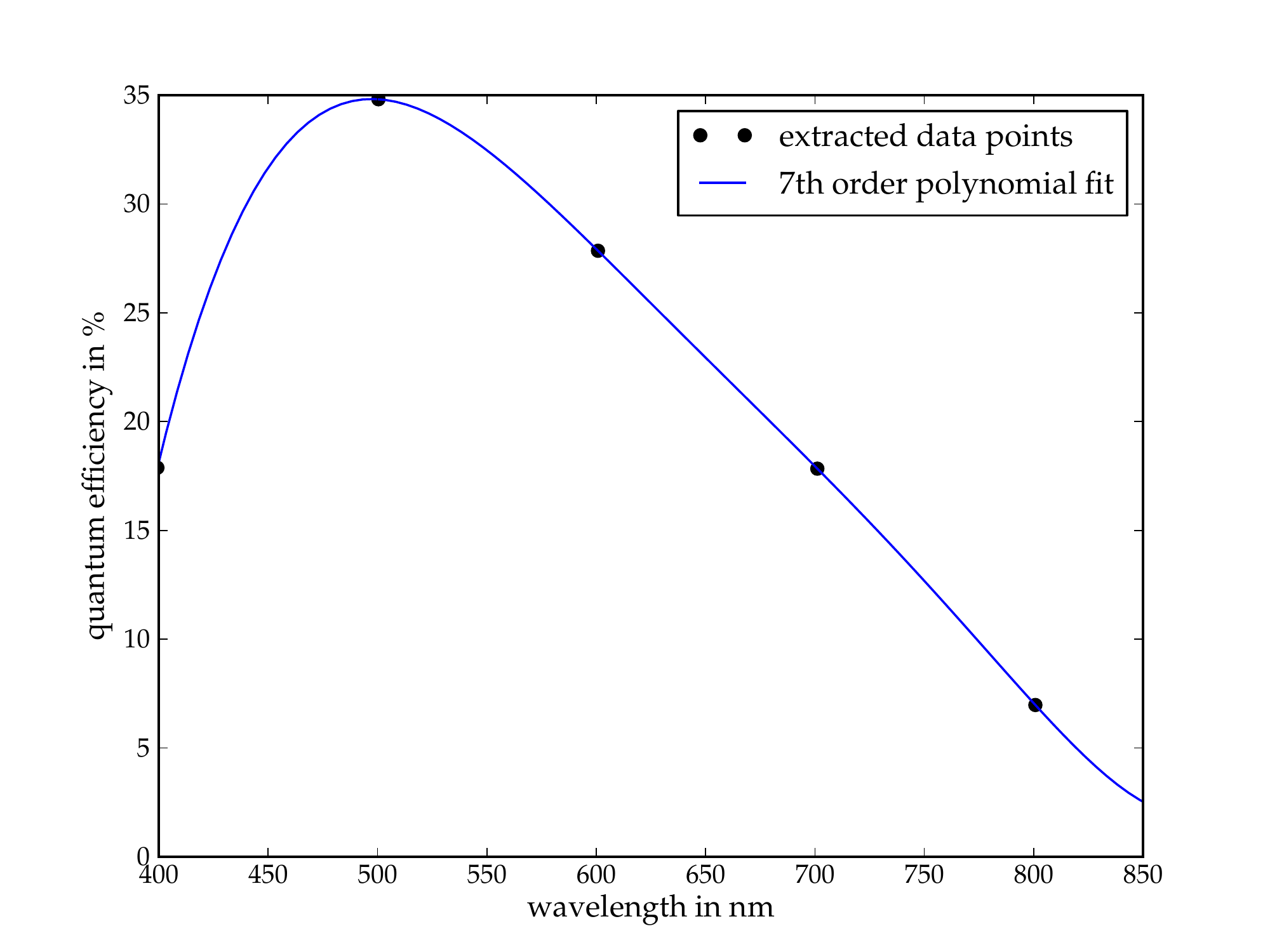}
\end{minipage}
\caption{Quantum efficiency 7th order polynomial fits for PicoQuant PDM series$^{1}$ (left) and the IDQuantique ID100 series$^{2}$ from figures on the manufacturers' data sheets using the WebPlotDigitizer$^{3}$ by Ankit Rohatgi.}
\label{fig:qefffit}  
\end{figure}
\begin{figure}
\begin{minipage}[b]{\linewidth}
\centering
\includegraphics[scale=.8]{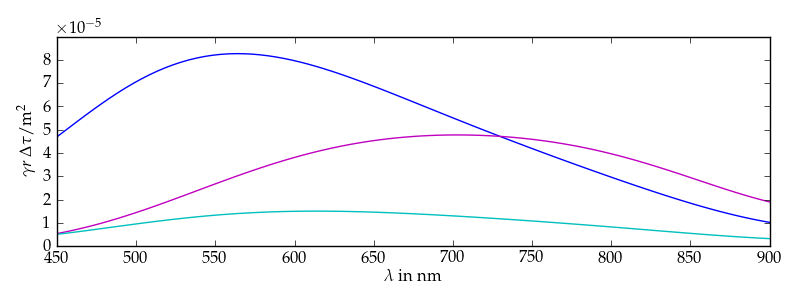} 
\end{minipage}
\begin{minipage}[b]{\linewidth}
\centering
\includegraphics[scale=.8]{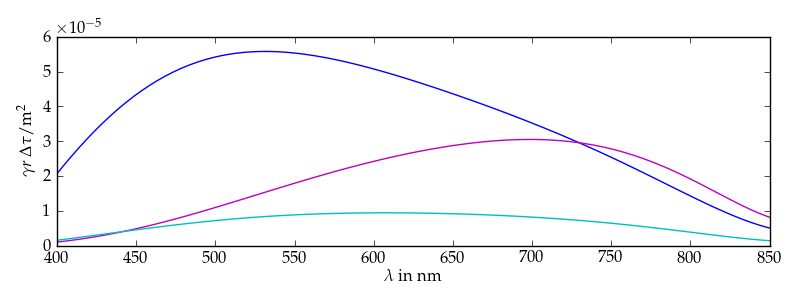} 
\end{minipage}
\caption{Upper panel: expected count rates per coherence time $\Delta\tau$ using two Picoquant PDM series photon counters after filtering to a bandwidth of $\Delta\lambda = 1\rm\,nm$ for different stars using the quantum efficiency estimates given in the left panel of figure \ref{fig:qefffit}. The plot shows equation (\ref{eq:snrarea2det}) evaluated for Sirius (blue), Betelqeuse (magenta), and Capella a (cyan) at the approximate parameters given in table \ref{stars}. The lower panel shows the expected count rates per coherence time for measurements using the IDQuantique ID100 series.}
\label{fig:qrate}
\end{figure}

For purposes of estimating count rates we approximate stars by black body discs. In particular we approximate Sirius (temperature 9940~K, diameter $0.006''$), Betelgeuse (3500~K, $0.04''$), and Capella~a (5700~K, $0.003''$). Let $r$ be the average count rate for one detector, the number of counts per coherence time $\Delta\tau$ then has the simple form \citep[][]{2014MNRAS.437..798M}:
\be
\gamma r\Delta\tau = \frac{\gamma A\Omega} {\lambda^2 (\exp{[h c/\lambda k_B T]} - 1)}
\label{eq:snrarea2det}
\ee
where $\Omega$ is the solid angle of the source, $A$ the collection area of the detectors, $\lambda$ the measurement wavelength, $T$ the surface temperature of the star, and $\gamma$ the quantum efficiency of the detector. Today's possibilities allow for reasonable quantum efficiencies at high time resolutions in the middle of the visible spectrum. 
Off the shelf photon counters such as the Pico Quant PDM series$^{[1]}$ or the ID100 series of IDQuantique$^{[2]}$ can help to get a feeling for achievable count rates (see best case parameters in Table \ref{detectors} and fits to quantum efficiency measurements by the manufacturers in Figure \ref{fig:qefffit}). 
Zooming in on the wavelength region that is covered by these detectors, Figure \ref{fig:qrate} shows the expected count rates for different stars using narrowband filters with $\Delta\lambda = 1\rm\,nm$.

\begin{table}
\centering
\begin{tabular}{l*{6}{c}r}
detector & max efficiency & wavelength of max efficiency & time resolution $\Delta t$& dead time \\
\hline
Picoquant & 49 \% & 550 nm & 50 ps & 80 ns \\
PDM series$^{1}$ &&&& \\
\hline
IDQuantique & 35 \% & 500 nm & 40 ps & 45 ns \\
ID100 series$^{2}$ \\
\hline
\end{tabular}
\caption{Some best case specifications of the two mentioned single photon counters.}
\label{detectors}
\end{table}
\footnotetext[1]{http://www.picoquant.com/images/uploads/downloads/pdm\_series.pdf} 
\footnotetext[2]{http://www.idquantique.com/images/stories/PDF/id100-single-photon-detector/id100-specs.pdf}
\footnotetext[3]{http://arohatgi.info/WebPlotDigitizer/app/}

\subsection{Signal to Noise Ratio for 2 Detectors}

Over a coherence time $\Delta \tau$ there will be $|V_{12}|^2(\gamma r\Delta \tau)^2$ coincidences. 
The time resolution, or counting time of a detector $\Delta t$ will be much larger than the coherence time, $\Delta t \gg \Delta\tau$, so in one counting interval there will be $\Delta t/\Delta\tau$ such contributions, giving $|V_{12}|^2 (\gamma r)^2\Delta\tau\Delta t$. Meanwhile there will be $(\gamma r\Delta t)^2$ random coincidences corresponding to the first term of equation (\ref{eq:3corr}),  giving $\gamma r\Delta t$ of noise, so the signal to noise ratio in $\Delta t$ becomes:
\be
SNR(\Delta t) =  |V_{12}|^2 \gamma r \Delta\tau ,
\ee
Note that the signal to noise ratio is independent on the bandwidth: decreasing bandwidth results in lower rates but will not change the signal to noise ratio.

\begin{figure}
\begin{minipage}[b]{\linewidth}
\centering
\includegraphics[scale=.8]{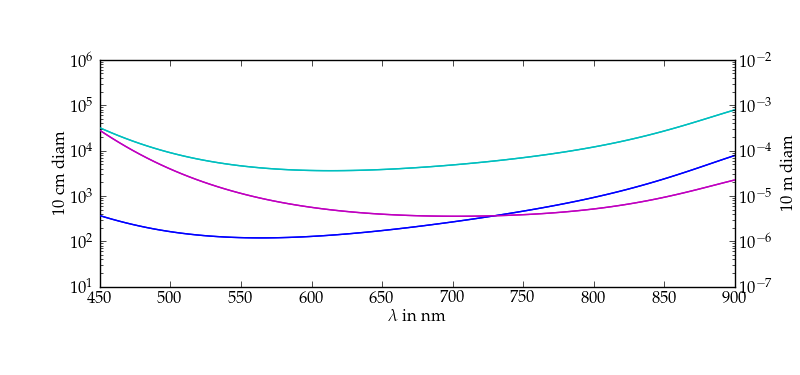} 
\end{minipage}
\vskip -1.4cm
\begin{minipage}[b]{\linewidth}
\centering
\includegraphics[scale=.8]{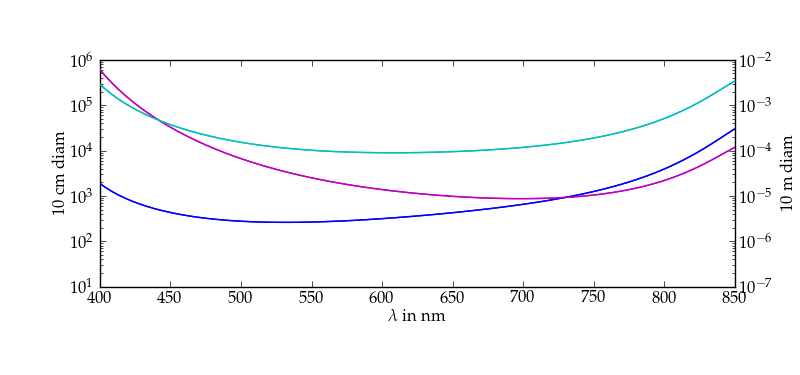} 
\end{minipage}
\caption{Upper panel: observation time estimates in seconds to reach an HBT SNR of $1$ for dishes with diameters of $.1\rm\,m$ (scales on the left) and $10\rm\,m$ (scales on the right) using the Pico Quant PDM series single photon counters, see table \ref{stars} for color code. The lower panel shows observation time estimates using ID100 detectors by IDQuantique.}
\label{fig:qtobs2}
\end{figure}

Evaluating the signal to noise ratio over an observation time $T_{\rm obs}$, we assume that the intensity fluctuations measured in one counting time $\Delta t$ are not correlated to the ones measured in the next interval, which is a reasonable assumption, since at relevant wavelengths $\Delta t\gg\Delta\tau\: (\sim 10^{-12}$). Thus, uncorrelated, Gaussian noise dominates and the SNR adds in quadrature:
\be
SNR = SNR(\Delta t) \sqrt{T_{\rm obs}/\Delta t}
\ee
We can now find the required observation time for a required signal to noise ratio using equation (\ref{eq:snrarea2det}):
\be
T_{\rm obs}(SNR) = \frac{SNR^2}{|V_{12}|^4}\Delta t  \Big[\frac{\gamma A \Omega} {\lambda^2 (\exp{[h c/\lambda k_B T]} - 1)}\Big]^{-2}
\label{eq:TOBS}
\ee
Figure \ref{fig:qtobs2} shows the the observation times necessary for some example cases to reach an SNR of $1$ for $|V_{12}|^2 = 1$. Clearly detector technology has improved greatly in the past half century, observing Sirius Hanbury Brown and Twiss needed a few minutes to reach a signal to noise ratio of 1 using a $5$-foot searchlight mirrors. Now the same could be done using $10\rm\,cm$ diameter mirrors.

\subsection{Signal to Noise Ratio for 3 Detectors}
\begin{figure}
\begin{minipage}[b]{\linewidth}
\centering
\includegraphics[scale=.8]{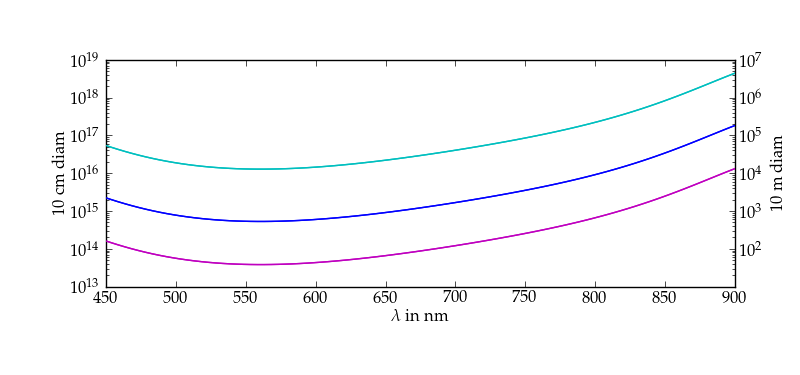} 
\end{minipage}
\vskip -1.4cm
\begin{minipage}[b]{\linewidth}
\centering
\includegraphics[scale=.8]{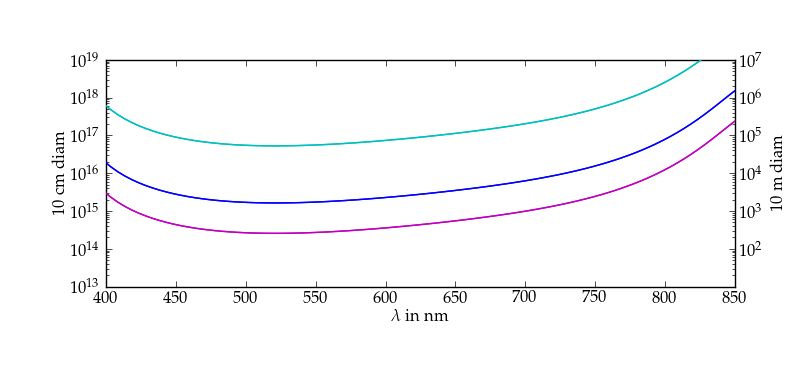} 
\end{minipage}
\caption{Upper panel: observation time estimates to reach an SNR of $1$ for $|V_{123}| = 10^{-3}$ for dishes with diameters of $.1\rm\,m$ (scales on the left) and $10\rm\,m$ (scales on the right) using Pico Quant PDM series single photon counters vs wavelength, see table \ref{stars} for color code. The lower panel shows observation time estimates using ID100.}
\label{fig:qtobs3}
\end{figure}

Over a coherence time $\Delta \tau$ there will be $V_{123}(\gamma r\Delta \tau)^3$ excess coincidences, where $V_{123}$ is the last term in equation (\ref{eq:3corr}). So over a time $\Delta t$ there will be $V_{123}(\gamma r)^3 \Delta \tau^2\Delta t$ signal coincidences. Meanwhile there will be $(\gamma r \Delta t)^{3/2}$ of noise, hence the signal to noise ratio is:
\be
SNR_{3}(\Delta t) \sim V_{123}(\gamma r\Delta\tau)^{3/2} (\Delta\tau/\Delta t)^{1/2} ,
\label{3detSNR}
\ee
again understanding $r\Delta \tau$ as the count rate at one detector over a coherence time. We can now rewrite and expand as before to find the observation time for a specific signal to noise ratio:
\be
T_{\rm obs}(SNR) = \frac{SNR^2}{V_{123}^2}\frac{\Delta t^2}{\Delta\tau}  \Big[\frac{\gamma A \Omega} {\lambda^2 (\exp{[h c/\lambda k_B T]} - 1)}\Big]^{-3}
\label{eq:SNRT3}
\ee
Comparing with equation (\ref{eq:TOBS}) we note the steeper dependence not only on the quantum efficiency $\gamma$ but also on counting time resolution $\Delta t$ and detector area $A$.
As we saw from Figures \ref{fig:simone} and \ref{fig:diffat}, features with brightness differences of around 25\% show up in $V_{123}$ at the level of $10^{-3}$. Figure \ref{fig:qtobs3} shows observation time estimates for our example cases for an $SNR$ of $1$ and $V_{123} = 10^{-3}$. We can immediately see that attempting three-point HBT with smaller sized telescopes is not an option but increasing the telescope diameters dramatically reduces observation times as the cube of the area thus making it possible to measure the bispectrum. With the equivalent of  $10\rm\,m$ diameter mirrors these should be easily detectable in the case of Betelgeuse or Sirius.

\section{Outlook}
The results shown in Figures \ref{fig:qtobs2} and \ref{fig:qtobs3} suggest that the recovery of HBT phase is feasible with present day detector technology and may lead to major advances in stellar imaging. Many technical problems will need to be solved first. The most important of these are the following.
\begin{itemize}
\item Designing suitable configurations for three detectors is essential. In particular, for Betelgeuse $10\rm\,m$ diameter mirrors would wash out the interference patterns. Some combination of small mirrors to resolve the large scale and large mirrors to resolve small scale structures is desirable.
\item An image reconstruction algorithm using two- and three-point HBT signal is needed.
\item Large mirrors such as proposed by CTA introduce non-isochronicity of a nanosecond or more \citep{2006ApJ...649..399L}.  In order to benefit from fast photon counting, the light paths from different parts of the mirror need to be equalized to the sub $\Delta t$ level (submillimeter) level.  Since only small fields of view are involved, it is plausible that a simple spherical-aberration compensators could do the job.
\end{itemize}

\section*{Acknowledgments}

We thank the referee, Paul Nu\~nez, for many helpful comments.

\clearpage
\bibliographystyle{astron}
\bibliography{paper}

\newpage

\begin{table}
\centering
\begin{tabular}{l*{6}{c}r}
source & temperature & diameter & color \\ 
& [K] & [arcsec] & \\
\hline
Sirius & 9940 & .006 & blue \\
\hline
Betelgeuse & 3500 & .04 & magenta \\
\hline
Capella a & 5700 & .003 & cyan \\
Capella b & 4940 & .004 &  \\
\hline
\end{tabular}
\caption{Approximated stellar parameters.}
\label{stars}
\end{table}

\end{document}